\documentstyle[aps,epsf]{revtex}
\begin{document}
\draft
 
\title{Pion Absorption Cross Section for $^{2}$H and $^3$He~
       in the $\Delta$-Isobar Region: A~Phenomenological Connection }

\author{
H.~Kamada$^1$,
M. P. Locher$^1$,
T.-S. H. Lee$^2$,
J.~Golak$^3$,
V. E. Markushin$^{1}$,
W.~Gl\"ockle$^4$,
H.~Wita\l a$^3$}

\address 
{$^1$Paul Scherrer Institut (PSI), CH-5232 Villigen, Switzerland;
\linebreak
$^2$Argonne National Laboratory, Argonne, Illinois 60439;
\linebreak
 $^3$Institute of Physics, Jagellonian University,
PL-30059 Cracow, Poland;
\linebreak 
 $^4$Institut f\"ur Theoretische Physik II, Ruhr Universit\"at Bochum,
 \linebreak D-44780 Bochum, Germany}

\date{\today}
 
\maketitle
\bigskip
\bigskip

\begin{abstract}
The absorption of $\pi^+$ on $^3$He in the $\Delta$-region is evaluated with
exact inclusion of the final state interaction among the three emerging
protons. The absorption is described by a $\pi N \rightarrow \Delta$ vertex
and  a $N\Delta$-$NN$ transition t-matrix which are calculated from a
phenomenological model for $NN$ and $\pi d$ reactions.
In a calculation where the initial pion scattering effects are neglected,
the predicted peaks of the pion absorption cross sections for $^2$H and
$^3$He  lie too high in energy in relation to the data. The effect of the
final state three-nucleon interaction turns out to be too small for
changing the magnitude and shifting the peak position of the total
absorption cross section for $^3$He.
We demonstrate that the adjustment of the peak position for the deuteron
cross section by small modifications of the $\Delta$-parameters,
automatically leads  to the correct peak position in $^3$He.
\end{abstract}

\bigskip
\bigskip

\section{Introduction}
\label{Intro}

In recent years it became possible to solve the quantum mechanical
three-body problem with  realistic two- and three-nucleon forces 
\cite{ref1,ref2,ref5,ref6,ref7}. Powerful computer facilities allowed this 
important step forward. Except for a few observables the theoretical
predictions  based on realistic $NN$ forces agree very well with the
experimental data. The exact treatment of the strong rescattering among the
three particles is thereby crucial.
 
 The Faddeev equations have been applied not only to the pure $3N$ system but
also to inelastic electron scattering on $^3$He 
\cite{ELECT1,ELECT2,ELECT3,ELECT4}. The Faddeev formalism allowed to
calculate any breakup process, exclusive \cite{ELECT3} and inclusive
\cite{ELECT4} ones.
In the same manner we apply now the Faddeev equations to investigate pion
absorption phenomena. The simplest reaction is pion absorption on the
deuteron requiring the study of the $\pi NN$ system for which  a vast
literature exists  
\cite{THEO7,THEO1,THEO2,THEO3,THEO4,THEO5,THEO6,LEE0,Lo,LOCHER,Sauer,AS86}. 
We shall not try to improve our understanding of this system, rather we
shall present an exploratory calculation of pion absorption on $^3$He which
is motivated by recent experimental studies of the reaction
${\pi^+}{^3}\mbox{He}\to3p$ by the LADS collaboration at PSI
\cite{LADS94,LADS1,LADS96}.
  In this first study we shall treat the dynamics of the incoming pion
approximately. In particular we shall not allow for initial state
interactions where the pion is rescattered before it is absorbed. We shall
assume that pion absorption takes place in the first step by a
$\Delta$-resonance mechanism and after that the nucleons interact strongly
in a $3N$ state.  
  We shall determine the effect of this final state interaction for the
total absorption cross section. To the best of our knowledge this will be
the first time that an exact treatment of FSI has been performed. The
choice of $^3$He instead of $^{2}$H also allows for pion absorption on 2
nucleons not only in isospin $t=0$ but also in $t=1$ states. Finally
choosing $\pi^+$ absorption on $^3$He generates a system of three
interacting final protons, which cannot be realized in a pure $3N$ scattering
process.

  In this study we are interested in the $\Delta$-resonance energy range and
therefore we introduce explicitly the $\Delta$ degree of freedom. We use the
phenomenological $NN$-$N\Delta$ model of Betz and Lee \cite{LEE0} which 
treats the $\pi N \Delta$ vertex that is reponsible for pion absorption
in a self consistent way. In the present exploratory calculation we
exclude for simplicity the contributions corresponding to propagating 
$\pi N N$ intermediate states. Ohta, Thies and Lee \cite{LEE} applied 
a similar simplification of the model of Betz and Lee to
heavier nuclei but did not include final state interactions and had
to rely on simple model target wavefunctions, whereas here we shall
use a realistic $^3$He description.
A glance at the pion absorption cross section for $^{2}$H and  $^3$He
reveals immediately that it peaks around $T _{\pi }\approx 130\;$MeV,
whereas the elastic and inelastic cross sections peak around 170~MeV,
closer to the position of the $\Delta$ resonance in free $\pi N$
scattering. We shall establish that FSI is not related to that shift 
in the peak position. However, it is possible to describe the energy 
shifts in both nuclei by a common parametrizatioon of the underlying 
mechanism.

In section~\ref{SecForm} we briefly outline the way we use the Faddeev
equations to describe the final state interaction for pion absorbtion on
$^3$He. Since it is similar in structure to inelastic electron scattering on
$^3$He we can refer to various  articles\cite{ELECT4} for more details and
show only those steps which are specific to the pion absorption process. 
This is presented in section~\ref{tech}. In order to show how we treat the
$\Delta$-particle we introduce the $N\Delta$ propagator in
section~\ref{SecDress}. A formalism very similar in structure to ours has
been presented before in \cite{THEO9}, though no numerical application
thereof is known to us.  Our numerical results are shown in
section~\ref{res}. We summarize and give an outlook in section~\ref{sum}.

\section{Formalism}
\label{SecForm} 

\begin{figure} \centering
\mbox{\epsfysize=60mm \epsffile{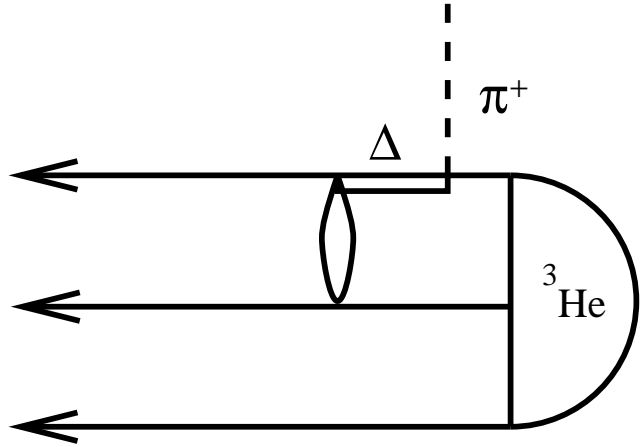}} \vspace*{4mm} 
\caption{The amplitude $\vert \Gamma\rangle$ of Eq.(\protect\ref{Lead})
  describing $\pi^{+} $ absorption on a nucleon in $^3$He leading to a
  $\Delta$-particle and followed by a deexcitation into two nucleons.}
\label{fey1}
\end{figure}

\begin{figure} \centering
\mbox{\epsfysize=60mm \epsffile{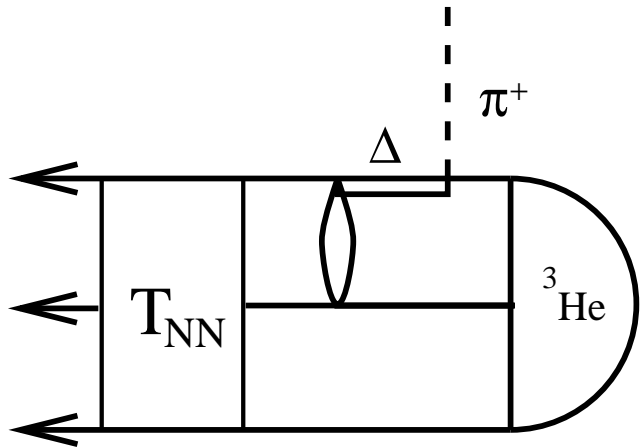}} \vspace*{4mm}
\caption{The $\pi$-absorption on $^3$He as described in
  Fig.\protect\ref{fey1} followed by the complete $3N$ final state
  interaction $T_{NN}$. }
\label{fey2}
\end{figure}

Let us first describe a situation where the pion is absorbed on a nucleon
converting it into a $\Delta$ particle which then together with a second
nucleon undergoes an infinite number of rescatterings described by a two-body
t-matrix $t_{NN,N\Delta}$. That two-body t-matrix obeys  a coupled set of
Lippmann-Schwinger equations
\begin{eqnarray}
 \left( \matrix{ t_{NN,NN} & t_{NN,N\Delta} \cr
                 t_{N\Delta, NN} & t_{N\Delta, N\Delta} \cr} \right)
 &=&
 \left( \matrix{ V_{NN,NN} & V_{NN,N\Delta} \cr
                 V_{N\Delta, NN} & V_{N\Delta, N\Delta} \cr} \right) \cr
 &+&
 \left( \matrix{ V_{NN,NN} & V_{NN,N\Delta} \cr
                 V_{N\Delta, NN} & V_{N\Delta, N\Delta} \cr} \right)
 \left( \matrix{ G_{NN}^{0} & 0 \cr 0 & G_{N\Delta}^{0} \cr} \right)
 \left( \matrix{ t_{NN,NN} & t_{NN,N\Delta} \cr
                 t_{N\Delta, NN} & t_{N\Delta, N\Delta} \cr} \right)
\label{LS}
\end{eqnarray}
In our model calculation we choose the transition potentials $V$ from the
analysis of Betz and Lee\cite{LEE0}. The iteration of Eq.(\ref{LS}) describes
the consecutive transitions between the $\Delta N$ system generated by the pion
absorption  and the resulting $NN$ system. The resulting amplitude has the
form
\begin{eqnarray}
  \vert \Gamma \rangle  = 
  t_{ NN, N\Delta } G^{0}_{NN\Delta} F(\pi) \vert \pi, ^3\mbox{He} \rangle
\label{Lead}
\end{eqnarray}
where $G^{0}_{NN\Delta}$ is the free
NN$\Delta$ propagator, $F(\pi)$ is the $\pi$-absorption vertex function and
$\vert\pi,^3\mbox{He}\rangle$ is the initial state. This term is depicted in
Fig.\ref{fey1}.  The amplitude $\Gamma$ is the starting point for the
rescattering processes among the three nucleons. Taken by itself it provides 
the properly symmetrised impulse approximation
\begin{eqnarray}
   U^{DWIA} = {1 \over \sqrt{3}} (1+P) \vert \Gamma \rangle
\end{eqnarray}
Here DWIA means distorted waves with respect to the two-body subsystem
and plane wave  with respect to the third particle.  We use the usual 
permutation operator $P$, a sum of a cyclic and an anticyclic permutation of
3 objects, which is a very convenient structural element in the Faddeev
treatment of three identical particles \cite{GTEXT}. The
three nucleon rescattering amplitude
\begin{eqnarray}
  U^{rescatt} = { 1 \over \sqrt{3}} (1+P) T_{NN}  \vert \Gamma  \rangle
\end{eqnarray}
is generated by the operator $T_{NN}$, which obeys
\begin{eqnarray}
  T_{NN} \vert\Gamma\rangle
  = t_{NN,NN} G^{0}_{NNN} P \vert\Gamma\rangle
  + t_{NN,NN} G^{0}_{NNN} P T_{NN} \vert\Gamma\rangle
\label{Faddeev1}
\end{eqnarray}
Here $T_{NN}$ is a three-body operator and $t_{NN,NN}$ is a two-body
operator as depicted in Fig.\ref{fey2}. Because of our simplifying
assumption a re-occurence of a $\Delta$-particle is not allowed, thus only
the free $3N$ propagator $G^{0}_{NNN}$ occurs.

Comparing  Eq.(\ref{Faddeev1}) to the  corresponding equation for inelastic
electron scattering \cite{ELECT4}, we see that the driving term is modified
due to the absence of the term $t_{NN,NN}G^{0}_{NNN}\vert\Gamma\rangle$.
That term would double-count the $NN$ interaction, since $\vert\Gamma\rangle $
contains the $NN$ interaction to inifinite order in the same particle channel.
 In electron
scattering  $\vert\Gamma\rangle$ is driven by the electromagnetic current
operator and no double counting occurs. 

For the processes discussed up to now the breakup amplitude is
\begin{eqnarray}
   U = U^{DWIA} + U^{rescatt} 
\label{UUU}
\end{eqnarray}
and this will be investigated numerically. 

\begin{figure} \centering
\mbox{\epsfysize=60mm \epsffile{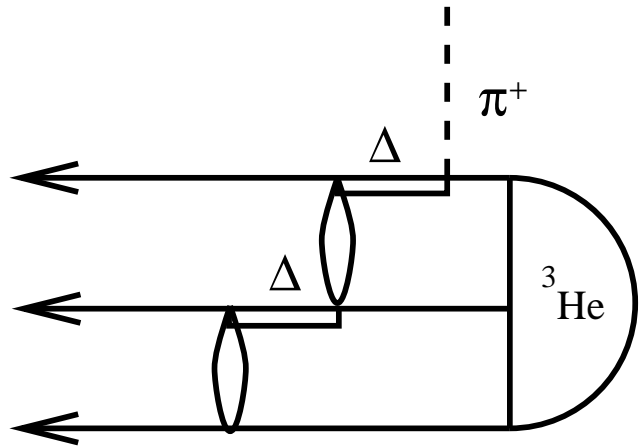}} \vspace*{4mm} 
\caption{Lowest order initial state interactions, ISI.}
\label{fey3}
\end{figure}

\begin{figure} \centering
\mbox{\epsfysize=60mm \epsffile{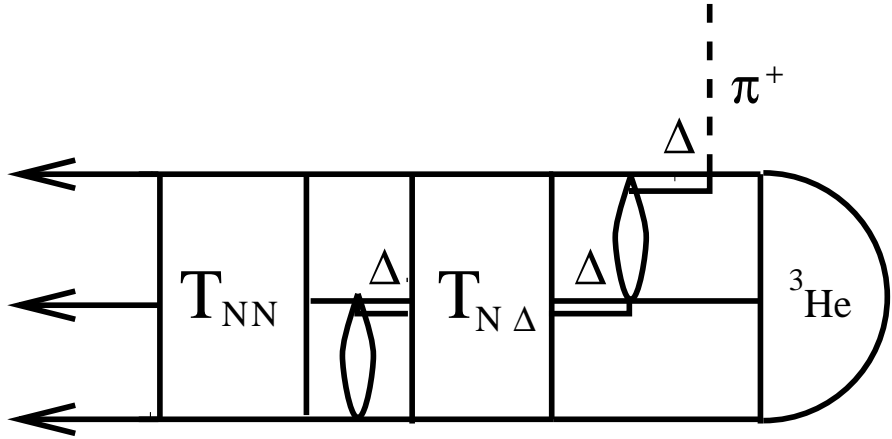}} \vspace*{4mm} 
\caption{The leading term of Eq.(\protect\ref{Faddeevfull}) representing
  initial state interactions acting in the Hilbert space of two nucleons and
  one $\Delta$-particle.}
\label{fey5}
\end{figure}

\begin{figure} \centering
\mbox{\epsfysize=60mm \epsffile{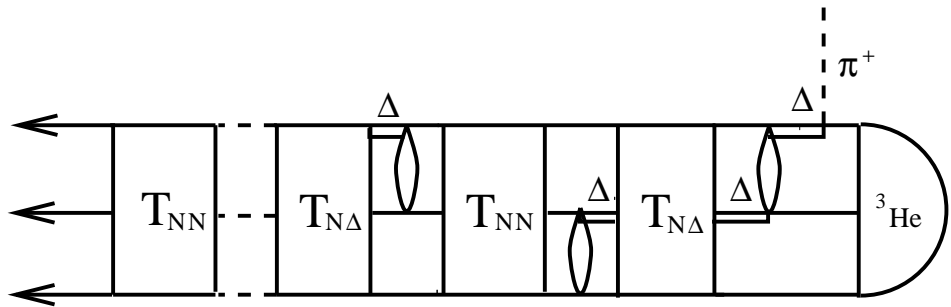}} \vspace*{4mm}
\caption{The general representation of the second term of
         Eq.(\protect\ref{Faddeevfull}).}
\label{fey6}
\end{figure}

So far the rescattering parts of the diagrams were three nucleon reducible.
Nonreducible diagrams shown in Fig.\ref{fey3} are   likely to play an
important role and will be investigated numerically in a forthcoming
article. Here we just present the necessary formal extensions. 
The $\Delta$ resulting from the the absorption of the initial pion can be 
absorbed and re-excited on another nucleon line, a process that can be 
iterated before the three nucleon final state is reached.
This is incorporated in the amplitude\footnote{In the $3N$ problem this
  quantity would be called $U_{0}$ with the index denoting the three
  nucleon continuum channel. This is unnecessary here since $\pi ^{+}$
  absorption on $^3$He has no other channels.}
\begin{eqnarray}
  U^{ISI} = {1 \over \sqrt{3}} (1 + P) T^{ISI} \vert\Gamma\rangle  
\label{UISI}
\end{eqnarray}
where the superscript $ISI$ stands for initial state interaction
and  $ T^{ISI}$ obeys the integral equation
\begin{eqnarray}
  T^{ISI} \vert\Gamma\rangle
  &=&
  T_{NN} \; t_{NN,N\Delta} G^{0}_{NN\Delta} P T_{N\Delta}
         F(\pi) \vert \pi, ^{3}\mbox{He} \rangle \cr
  &+&
  T_{NN} \; t_{NN,N\Delta} G^{0}_{NN\Delta} P T_{N\Delta} \; 
         t_{N\Delta,NN} G^{0}_{NNN} P T^{ISI} \vert\Gamma\rangle  
\label{Faddeevfull}
\end{eqnarray}
and $T_{N\Delta}$ generates all possible $N\Delta$ pairs  via 
\begin{eqnarray}
  T_{N\Delta}  = t_{N\Delta,N\Delta} G^{0}_{NN\Delta}
               + t_{N\Delta,N\Delta} G^{0}_{NN\Delta} P  T_{N\Delta} \ . 
\label{Faddeev2} 
\end{eqnarray}
Iterating Eqs.(\ref{Faddeev1},\ref{Faddeevfull},\ref{Faddeev2}) one can
visualize the processes contained in $T^{ISI}$ as is shown in
Figs.\ref{fey5},\ref{fey6}. Fig.~\ref{fey3} is the simplest new diagram
contained in Fig.\ref{fey5} representing the initial state interaction. An
example of an additional final state interaction (the leading term of
Eq.(\ref{Faddeevfull})) is shown in Fig.\ref{fey5}.

\section{Choice of Coordinates}
\label{tech}

The amplitude $|\Gamma\rangle$ contains three steps: the pion absorption 
by the single particle operator $F(\pi)$, the free propagator of the (zero
width) $\Delta$-particle and two nucleons,  and the action of the transition
operator $t_{NN,N\Delta} $ converting the $N\Delta$ system into a
two-nucleon system. In our three-body context this requires the use of
various sets of Jacobi momenta. The $^3$He wavefunction depends on the
following momenta
\begin{eqnarray}
  {\vec p}^{\ '} & = & \frac12 ( {\vec k}_2^{\ '} - {\vec k}_3^{\ '})
  \nonumber \\
  {\vec q}^{\ '} & = & \frac23 ({\vec k}_1^{\ '} - \frac12 ( {\vec k}_2^{\ '}
                 + {\vec k}_3^{\ '} ) )
\label{jacprime}
\end{eqnarray}
defined in terms of the individual momenta of three nucleons. After the pion
absorption on nucleon~1 we describe the system consisting of two nucleons
and a $\Delta$-particle by
\begin{eqnarray}
  {\vec p} & = & \frac12 ( {\vec k}_2 - {\vec k}_3 ) \nonumber \\
  {\vec q} & = & \frac{2 M_N {\vec k}_1 - (M_N + \omega) ({\vec k}_2 +
                 {\vec k}_3 )} {3 M_N + \omega} \ \ ,
\label{jacnoprime}
\end{eqnarray}
where $M_N$ is the nucleon mass and $\omega=\sqrt{\mu^2+k_\pi^2}$ the energy
of the pion\footnote{Following Ref.\cite{LEE0}
the quantity $M_N + \omega$ is used intechangeably with
$M_\Delta$ in the resonance energy range.}  in the overall CMS.
We choose the single particle
operator  $F(\pi)$  to depend on the relative momentum ${\vec q}_0 $ of
nucleon~1 and the pion:
\begin{equation}
  {\vec q}_0 = \frac{ M_N {\vec k}_\pi - \omega {\vec k}_1^{\ '} }
               {M_N + \omega} =
  \vec k _{\pi } -  \frac{ \omega } {M_N + \omega} \vec q   \ \ .
\label{q0}
\end{equation}
The second equality holds true in the overall CMS. 
The functional dependence of $F(\pi)$ related to the p-wave property of the
$\Delta$-particle is given in the Appendix. We define
\begin{equation}
  \langle {\vec k}_1 \mid F \mid {\vec k}_1^{\ '} {\vec k}_\pi \rangle
  = F({\vec q}_0) \; \delta({\vec k}_1 - {\vec k}_1^{\ '} - {\vec k}_\pi)  \ \
\label{F}
\end{equation}
where ${\vec k}_1$ is the momentum of the $\Delta$-particle. Therefore we
have 
\begin{eqnarray}
  \langle {\vec p}{\vec q} |F| \pi, ^3\mbox{He} \rangle
  & = & \int d{\vec p}^{\ '} \, d{\vec q}^{\ '} \,
  \langle {\vec p}{\vec q} |F| {\vec p}^{\ '}{\vec q} ^{\ '} \rangle \;
  \langle {\vec p}^{\ '}{\vec q}^{\ '} |\pi, ^3\mbox{He} \rangle
  = \nonumber \\  & = & 
  \int d{\vec q}^{\ '} \, F({\vec q}_0) \,
  \delta({\vec q} - {\vec q}^{\ '} - \frac23 {\vec k}_\pi) \,
  \langle {\vec p}{\vec q}^{\ '} |\pi, ^3\mbox{He} \rangle \ \ .
\label{F2}
\end{eqnarray}

The transition operator $t_{NN,N\Delta}$ acting between particles 1 and 2 
requires another set of Jacobi momenta
\begin{eqnarray}
  {\vec p}^{\ ''} & = & \frac{ M_N {\vec k}_1 - (M_N + \omega) {\vec k}_2 }
                             {2 M_N + \omega} \label{jacdbprimeA} \\
  {\vec q}^{\ ''} & = & \frac{ (2 M_N + \omega ) {\vec k}_3 - M_N
                        ({\vec k}_1 + {\vec k}_2)} {3 M_N + \omega}   \ \ .
\label{jacdbprimeB}
\end{eqnarray}
They are related to ${\vec p}$ and ${\vec q}$ by
\begin{eqnarray}
  {\vec p} & = & - \frac{3M_N + \omega}{2(2M_N + \omega)}
                   {\vec q}^{\ ''}  - \frac12 {\vec p}^{\ ''} \\
  {\vec q} & = & - \frac{M_N + \omega}{2M_N + \omega} {\vec q}^{\ ''}
                 + {\vec p}^{\ ''}   \ \ .
\label{jacrel}
\end{eqnarray}

Finally we use Jacobi momenta denoted by ${\vec p}^{\ '''}$ and
${\vec q}^{\ '''}$ for three nucleons analogous to
(\ref{jacdbprimeA},\ref{jacdbprimeB}) describing the three-nucleon system
to the left of the $NN,N \Delta$ transition matrix, see Fig.\ref{fey2}.
Since the transition $t$-matrix is diagonal in ${\vec q}^{\ '''}$ and
${\vec q}^{\ ''}$ we finally get 
\begin{eqnarray}
  \langle {\vec p}^{\ '''}{\vec q}^{\ '''} |
  t_{NN,N\Delta} G^{0}_{NN \Delta} F |\pi, ^3\mbox{He} \rangle
  = \hspace*{60mm} \nonumber \\
  \int d{\vec p}^{\ ''} \,
  \langle {\vec p}^{\ '''} | t_{NN,N\Delta} | {\vec p}^{\ ''} \rangle
  G^{0}_{NN \Delta} ({\vec p}^{\ ''}, {\vec q}^{\ '''}) 
  \int d{\vec p} \, d{\vec q} \,
  \langle {\vec p}^{\ ''} {\vec q}^{\ '''} | {\vec p}{\vec q} \rangle \;
  \langle {\vec p} {\vec q} |F|  \pi, ^{3}\mbox{He} \rangle  \ \ .
\label{finget}
\end{eqnarray}

\section{Dressing the $\Delta$-particle }
\label{SecDress}

So far we have introduced the momentum space representation of the
$NN-N\Delta$ transition  operator and the free $N\Delta$ propagator. In the
Betz-Lee model \cite{LEE0} the $N\Delta$ propagator is dressed 
\begin{eqnarray}
  G^0_{N\Delta} & = & 
  \frac{1}{E - (M_\Delta^0 - M_N) -
  \frac{\textstyle k^2 (M_{N}+M_{\Delta}^0)}{\textstyle 2M_{N}M_{\Delta}^0} - 
  \Sigma_{N\Delta}(k,E)} \ \ ,
\label{gnd}
\end{eqnarray}
where $E$ is the CMS energy of the two-nucleon system, $\vec k$ is
the $N\Delta$ relative momentum. The physical mass of the $\Delta$-particle is 
\begin{eqnarray}
   M_\Delta = M^{0}_{\Delta} + \delta M 
\end{eqnarray}
where $M_{\Delta}^0$ is the bare mass. 
The energy dependent self interaction $ \Sigma $ is
\begin{eqnarray}
 \Sigma_{N\Delta} (k,E) & = &  
 \int_0^\infty \, { {F^2(k') {k'}^{\ 2} dk'} \over
       {E + i\epsilon - H_{NN\pi}^0(k,k') - H_{N\pi}^0(k')} } 
 = \delta M - i \Gamma / 2 
\label{sigma}
\end{eqnarray}
with 
\begin{eqnarray}
   H_{NN\pi}^0 ( k,k') & = &  \frac{k^2}{2 M_N} +
                \frac{k^2}{2(M_N + \sqrt{\mu^2 + {k'}^{\ 2}})} \ \ , \\  
   H_{N\pi}^0 (k') & = &
                \frac{{k'}^{\ 2}}{2 M_N} + \sqrt{\mu^2 + {k'}^{\ 2}} \ \ .
\end{eqnarray}
Here $\Gamma $ is the energy dependent width of the $\Delta $-particle. The
vertex function $F$ contains the bare coupling constant  $F_\Delta^0$ and
the range parameter  $\Lambda_\Delta$,  which are defined in the Appendix.
The steps required for the partial wave representation are also described
there. For the calculation below we shall allow small variations
of the bare $\Delta$ parameters $M_{\Delta}^0$ and $F_{\Delta}^0$.

\section{Results}
\label{res}

In order to test the input for the pion absorption reaction on $^3$He we 
recalculated  the total pion  $\pi d$ absorption cross section on the
deuteron as a function of energy in the Betz-Lee model.  Table~\ref{TAB1}
shows the partial waves used, and the parameters for the potential are taken
from \cite{LEE0}. 

\begin{figure} \centering
\mbox{\epsfysize=8cm \epsffile{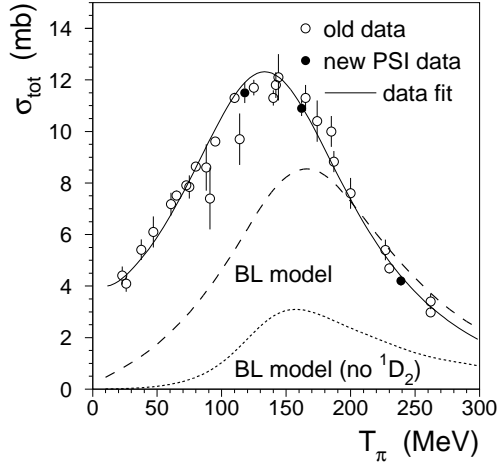}} \vspace*{4mm}
\caption{Total cross section for $\pi+d\to pp$ as a function of the
  laboratory pion kinetic energy. The data (solid line) are taken from
\protect\cite{Ritchie,piDexp0,piDexp1,piDexp2,piDexp3,piDexp4,piDexp5,piDexp6,piDexp7,piDexp8,piDexp9,piDexp10,piDexp11}.
  The dashed and dotted lines are calculated from the Betz-Lee
  potential with all partial wave set of Table \protect\ref{TAB1}
 and without the $^1D_2$ wave, respectively.} 
\label{f2}
\end{figure}

In the present study we exclude for simplicity the contribution corresponding
to propagating $\pi NN$ states and the initial pion scattering effects. Thus,
our result for the deuteron is different from the full unitary calculation of
Ref.[17].
The dashed line in Fig.\ref{f2} shows the total cross section together with
the data interpolated by the solid line. The dotted line shows a calculation
without the $^1D_2$ partial wave, which demonstrates the importance of that
wave.

\begin{figure} \centering
\mbox{\epsfysize=8cm \epsffile{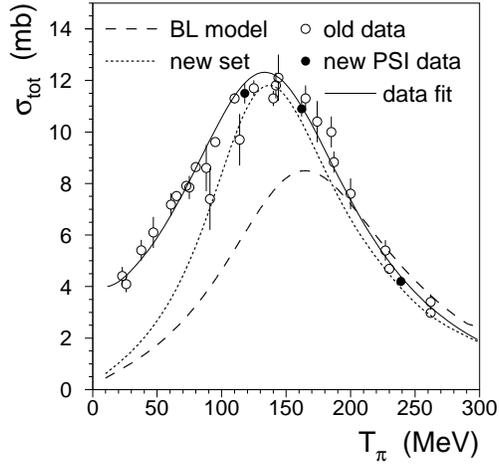}} \vspace*{4mm} 
\caption{Total cross section for $\pi+d\to pp$ as a function of the incident
  pion energy in laboratory system. The description of the lines is the same
  as in Fig.\protect\ref{f2} with the exception of the dotted line
  which is now the theoretical prediction based on the new parameter set.}
\label{f3}
\end{figure}

As discussed in Ref.[17], the Betz-Lee model in our simplified approximation
does not quantitatively reproduce the data.
The peak position is about 30 MeV too
high and the cross section is too low on the rising part below the
resonance. The last feature is certainly partly related to neglecting
non-resonant $\pi N $ partial waves. On the other hand, the shift of the
resonance has been obtained correctly in models containing  explicit pion
propagaton in intermediate states\cite{LEE0,Lo}. In the present paper we
stick to the pure $\Delta$-model excluding  explicit $NN\pi$ propagation for
the reaction on the deuteron or $NNN\pi $ in the case of $^3$He. We have
therefore adjusted  the bare parameters $M^{0}_{\Delta}$ and
$F^{0}_{\Delta}$ of the Betz-Lee model in order to reproduce the observed 
energy dependence of the total cross section on the deuteron. Only small
changes are needed to reproduce size and position at the resonance, see Fig
\ref{f3}:
\begin{eqnarray}
M_{\Delta}^{0} = 1280 \to 1260 \;\mbox{\rm MeV} 
\end{eqnarray}
and 
\begin{eqnarray}
F_{\Delta}^{0} = 0.98 \to 1.00 
\end{eqnarray}

\begin{figure} \centering
\mbox{\epsfysize=8cm \epsffile{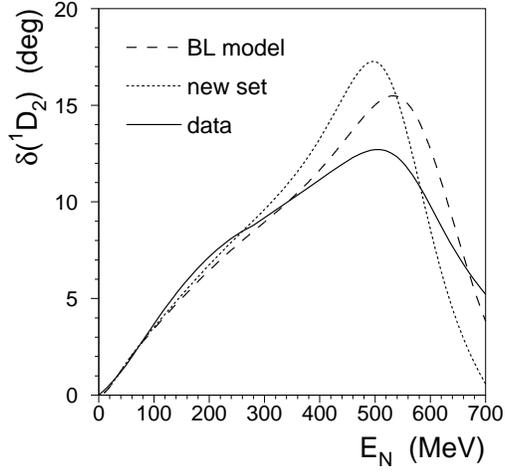}} \vspace*{4mm} 
\caption{The $^1D_2$ $NN$ phase shift as a function of lab. kinetic
         nucleon energy. The data (solid line) are represented by the
         partial wave from the SAID  analysis \protect\cite{SAID}. 
    The dashed and dotted lines correspond to the Betz-Lee potential
    and its new parameterization, respectively.} 
\label{f4}
\end{figure}

Since the new parameters reflect pion propagation in the absorption reaction in
an effective way, it is clear that the elastic $\pi N $ and the elastic $\pi d$
cross sections will not be correctly described. The same is true for the $NN$
phase shifts. As an illustration we show the effect of the new parametrization
on the $^{1}D_{2}$ partial wave in Fig.\ref{f4}. The Betz-Lee approach which
we use also
neglects the diagonal potential $V_{N\Delta, N\Delta}$. We have verified that
the inclusion of such  a potential allows to shift the peak position
downwards. For the time being, however, we restrict ourselves to treating
$M_{\Delta}^0$ and $F_{\Delta}^0$ as the only effective parameters and
retain $V_{N\Delta,N\Delta}$=0. 

\begin{figure} \centering
\mbox{\epsfysize=8cm \epsffile{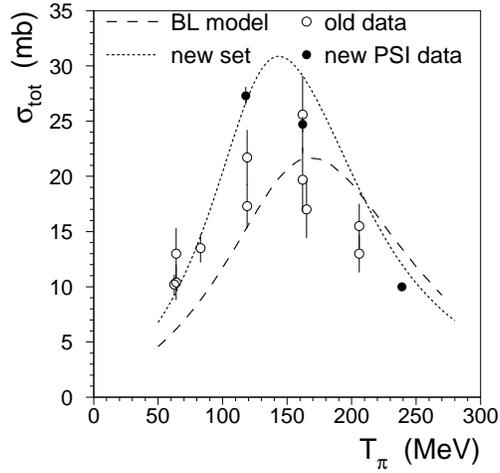}} \vspace*{4mm} 
\caption{Total cross section for $\pi^+ {^3}\mbox{He}\to ppp$ as a function of
     the laboratory pion kinetic energy. 
     The data are from PSI \protect\cite{LADS94,THESIS} and Refs.
\protect\cite{piHexp2,piHexp3,piHexp4,piHexp5,piHexp6,piHexp7,piHexp8}.
}
\label{f1}
\end{figure}

It is gratifying that the effective parameters of Fig.\ref{f3} also improve
the description of the total cross section on $^3$He as is shown in Fig
\ref{f1}. We therefore see that the gross feature on $^3$He falls into place
once the reaction on the deuteron is properly described.
In Fig.\ref{f1} the effects of final state interactions in the $3N$ continuum
state are fully included for the first time. On the scale of the figure the
effect of FSI is too small to be drawn (2\%). For the total cross 
section FSI is
thus  negligible. For observables and kinematics which are not dominated by
the two nucleon DWIA mechanism  (the quasi-deuteron process) significant
modifications due to FSI are however to be expected. 

\section{Summary and Outlook}
\label{sum}

We formulated a model of pion absorption on $^3$He in a Faddeev scheme, 
which includes  the final state interaction among the three outgoing
nucleons and which also allows for initial state interaction where more
than one $\Delta$-resonance is excited (see Fig.\ref{fey3}-\ref{fey6}).
The numerical
evaluation in this paper is restricted to the leading quasi-deuteron
absorption term with inclusion of the final state interactions, 
Eq.(\ref{UUU}), between the three protons. 
The phenomenological Betz-Lee model for the $NN$-$NN$ and  $NN$-$N\Delta$
systems is used. In the present exploratory calculation
where the initial pion scattering and the contributions 
corresponding to propagating $\pi NN$ states are neglected,
the resulting total pion absorption cross sections for
$^2$H and $^3$He do not agree with the data. The most striking feature is,
that the theoretical peak positions are  too high in comparison to
experiment.  The full inclusion  of the final state interaction among the 3
nucleons in our model has no visible  effect for the total cross section,
its contribution is only about 2\%. 

We introduced a very simple method to shift the peak position by treating
the  bare $\Delta$-mass $M^{0}_{\Delta}$ and the coupling strength
$F^{0}_{\Delta}$ as parameters. We demonstrated, that a slight modification
of order 2\% reproduces the  $^{2}$H and $^3$He cross sections at the same
time.  This points to a common mechanism for pion absorption in both nuclei. 

The Betz-Lee model sets the transition potential $V_{N\Delta,N\Delta}$ to zero.
In a forthcoming study we shall abolish that assumption and include the 
transition potential $V_{N\Delta,N\Delta}$, as it occurs
for instance in the phenomenological AV28 potential \cite{AV28}. 
At the same time we shall investigate the importance of initial state 
interactions introduced in Eq.(7). The diagonal $N\Delta$-potential
is expected to be important in this context.


\section*{Acknowledgements}

We thank C.H.Q. Ingram, R.P. Redwine, A. Lehmann, A.O. Mateos and N.K.
Gregory for valuable  discussions. The computational work was performed on
the NEC SX3 at the Swiss Center for Scientific Computing in Manno and on the 
CRAY J90 at the Eidgen\"ossische Technische Hochschule Z\"urich. 
This work was supported in part by the U.S. Department of Energy,
Nuclear Physics Division, under Contract No. W-31-109-ENG-38.

\section*{Appendix: Partial wave representations}

\begin{figure} \centering
\mbox{\epsfysize=60mm \epsffile{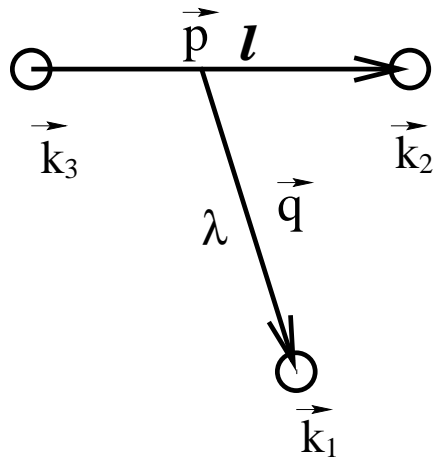}} \vspace*{4mm} 
\caption{Jacobi momenta and related orbital angular momenta.}
\label{jacobi}
\end{figure}

Here we present the partial wave representation used. It is related to the
choice of Jacobi coordinates of Fig.\ref{jacobi}, see also Sec.\ref{tech}. For
three particles (three nucleons or two nucleons and a $\Delta$-particle)
the partial wave basis in momentum space is 
\begin{eqnarray} 
\vert  p,  q, \alpha JMTM_T \rangle =
 \vert  p, q, ( s_2 s_3 ) s ( l s ) j ( \lambda s_1) I ( j I) J M 
( \tau_2 \tau_3) t ( t \tau_1 ) T M_T \rangle 
\label{PARTIAL}
\end{eqnarray} 
where the orbital angular momenta  $l$ and $\lambda$ are related to $\vec p$
and $ \vec  
q$ , 
and $s_i , \tau_i $ ( i=1,2,3) are spins and isospins, respectively. 

The initial state nucleus, the $^3$He ground state, has  $J={1\over2}$ and
$T={1\over2}$. The final $ppp$ state has $T={3\over2}$.
The $\pi N\Delta$ vertex function is written as 
\begin{eqnarray}
  F(\vec q_0) = F (q_0)
  \sum_{m,\mu} \vert \mbox{$3\over2$} \mu \mbox{$3\over2$} m \rangle
  \sum_{m_1,m_N} \langle \mbox{$3\over2$} m \vert
                 1 m_1 \mbox{$1\over2$} m_N \rangle
  Y^*_{1,m_1}(\hat q_0) \cr
  \langle \mbox{$3\over2$} \mu \vert 1 \mu_\pi \mbox{$1\over2$} \mu_N \rangle
  \langle \mbox{$1\over2$} m_N \mbox{$1\over2$} \mu_N \vert 
\end{eqnarray}
where $m$ and $m_N$ are the z-component of the spin of the $\Delta$ and the
nucleon, $\mu$ and $\mu_N$ are the corresponding isospin quantum numbers.
The pion enters through its orbital angular momentum z-component $m_1$ and
its isospin quantum number $\mu_{\pi}$.

The function $F(q_0)$ is taken from the  Betz-Lee model \cite{LEE0}
(Eq.(3.5)):
\begin{eqnarray}
   F(q_0)= {F_\Delta^{0} \over \sqrt{ 2 ( M_N + \mu )}}
           {q_0 \over \mu} 
   \left( {{\Lambda_\Delta^2} \over {\Lambda_\Delta^2 + q_0^2}} \right)^2
\end{eqnarray}
where $F_\Delta^0=0.98$ and $\Lambda_\Delta=358\;$MeV/c.
Using this operator $F$ of Eq.(\ref{F2}) can be written as 
\begin{eqnarray}
   \langle p,q,\alpha'J'M'T'M_T' \vert F \Psi J M T M_T \rangle =
   \sum_\alpha \delta_{l,l'} \delta_{s,s'} \delta_{j,j'} \delta_{t,t'} 
   (X_1 + X_2) {\cal I}
\end{eqnarray}
where 
\begin{eqnarray}
  X_1 = k_\pi \sqrt{3 \over { 4 \pi }}
      (-)^{\lambda+J'+I+j-J-M}
      \sqrt{\hat\lambda \hat\lambda' \hat I \hat I' \hat J \hat J'}
      \sum_{\lambda_1+\lambda_2=\lambda}
      q^{\lambda_1} ({2\over3} k_\pi)^{\lambda_2}
      \sqrt{{\hat\lambda !} \over {\hat\lambda_1 ! \hat\lambda_2 !}}
      \sqrt{\hat\lambda_1 \hat\lambda_2}
      \nonumber \\
      \times \sum_{b'} \hat b' \sum_{\cal L} \hat{\cal L}
      \left \{ \matrix{\lambda_1 & \lambda_2 & \lambda \cr
                              b' & \lambda ' & {\cal L} } \right \}
      \left( \matrix{{\cal L}& \lambda_1 & \lambda ' \cr 0 & 0 & 0 } \right)
      \left( \matrix{{\cal L } & \lambda_2 & b' \cr 0 & 0 & 0 } \right)
      S^\alpha_{\cal L}(p,q,k_\pi)
      \nonumber \\
      \times \sum_{x} (-)^x \hat x
      \left \{ \matrix{ I' & I & x \cr \lambda' & \lambda & b' \cr 
             {3\over2} & {1\over2} & 1 } \right \}
      \left(\matrix{x &  b' &  1 \cr 0 & 0 & 0 } \right)
      \left\{ \matrix{ J & I & j \cr I' & J' & x } \right\}
      \left( \matrix{J & x & J' \cr M & 0 & -M} \right) \ \ ,
\end{eqnarray}
\begin{eqnarray}
   X_2 = -\epsilon q \sqrt{3\over{4\pi}} (-)^{J'-J+j+I'-M}
   \sqrt{\hat J \hat J' \hat I \hat I' \hat\lambda \hat\lambda'}
   \sum_{\lambda_1+\lambda_1=\lambda}
   q^{\lambda_1} ({2\over3} k_\pi)^{\lambda_2}
   \sqrt{{\hat\lambda !}\over{\hat\lambda_1 ! \hat\lambda _2 !}}
   \sqrt{\hat\lambda_1 \hat\lambda_2}
   \nonumber \\
   \times \sum_{b b'} \hat b \hat b ' \sum_{\cal L} \hat{\cal L}
   \left \{ \matrix{\lambda_1 & \lambda_2 & \lambda \cr
          b' & b & {\cal L} } \right\}
   \left( \matrix{{\cal L} & \lambda _1 & b  \cr 0 & 0 & 0 } \right)
   \left(\matrix{{\cal L} &  \lambda _2 & b' \cr 0 & 0 & 0 } \right)
   S^\alpha_{\cal L} ( p,q,k_\pi)
   \nonumber \\ 
   \times \left( \matrix{\lambda' & 1 & b \cr 0 & 0 & 0 } \right)
   \left\{ \matrix{ I' & I & b' \cr \lambda & b & {1\over2}} \right \}
   \left\{ \matrix{ {3\over2} & {1\over2} & 1 \cr b & \lambda' & I'} \right\}
   \left \{ \matrix{ J & I & j \cr I' & J' & b'} \right\}
   \left( \matrix{J &  b'&  J' \cr M & 0 & -M} \right) \ \ , 
\end{eqnarray}
\begin{eqnarray}
  {\cal I} = \sqrt{4 \hat T \hat T '} (-)^{{1\over2}+t+T'-T-M_T'}
  \left\{ \matrix{ T & {1 \over 2 } & t \cr {3\over2} & T ' & 1} \right\}
  \left( \matrix{ T & 1 & T ' \cr M_T & \mu_\pi  & -M_T '} \right)
\end{eqnarray}
and 
\begin{eqnarray}
     \epsilon = {\omega\over{M_N+\omega}} 
\end{eqnarray}
The function $S_{\cal L}^\alpha(p,q,k_\pi)$ is defined as
\begin{eqnarray}
  S_{\cal L}^{\alpha }(p,q,k_\pi)
  = \int_{-1}^1 dx P_{\cal L}(x)
  { {F(\vert\vec k _\pi - \epsilon \vec q \vert)}
  \over {\vert\vec k _\pi - \epsilon \vec q \vert } }
  { {\Psi_{\alpha}(p,\vert\vec q-{2\over3}\vec k_\pi\vert)}
  \over {{\vert\vec q-{2\over3}\vec k_\pi\vert^ \lambda}} }
\end{eqnarray}
where $x$ is the cosine between $\vec q$ and $\vec k_\pi$,
and $\Psi_{\alpha}$ is the $^3$He wave function 
in the  basis (\ref{PARTIAL}).




\narrowtext
\begin{table}
\caption{Partial wave decomposition of $NN$ and $N\Delta$ systems.}
\label{TAB1}
\begin{tabular}{rcl}
\hspace*{20mm} $NN$  & \hspace*{10mm} & $N\Delta$ \hspace*{20mm} \\
\hline
   $^1S_0$           & &  $^5D_0$ \\
   $^3P_0$           & &  $^3P_0$ \\
   $^3P_1$           & &  $^3P_1$, $^5P_1$ \\
   $^3P_2$, $^3F_2$  & &  $^3P_2$, $^5P_2$ \\
   $^1D_2$           & &  $^5S_2$ \\
   $^3F_3$           & &  $^5P_3$ \\
   $^1G_4$           & &  $^5D_4$ \\
\end{tabular}
\end{table}

\widetext
\clearpage

\end{document}